\documentclass[a4paper, 10pt, conference]{IEEEtran}
\IEEEoverridecommandlockouts
\bstctlcite{IEEEexample:BSTcontrol}

\usepackage{cite}
\usepackage{amsmath,amssymb,amsfonts}
\usepackage{algorithmic}
\usepackage{graphicx}
\usepackage{textcomp}
\usepackage{xcolor}
\def\BibTeX{{\rm B\kern-.05em{\sc i\kern-.025em b}\kern-.08em
    T\kern-.1667em\lower.7ex\hbox{E}\kern-.125emX}}
\begin{document}

\title{Effect of different splitting criteria on the performance of speech
emotion recognition}
%

\author{
\IEEEauthorblockN{Bagus Tris Atmaja\textsuperscript{*}\thanks{*Corresponding author, on leave from Department of Engineering Physics, Institut Teknologi Sepuluh Nopember, Surabaya, Indonesia (email: bagus@ep.its.ac.id).}}
\IEEEauthorblockA{\textit{National Institute of Advanced Industrial Science and Technology, Japan}\\
b-atmaja@aist.go.jp}
\\
\IEEEauthorblockN{Akira Sasou}
\IEEEauthorblockA{\textit{National Institute of Advanced Industrial Science and Technology, Japan} \\
a-sasou@aist.go.jp}}

\maketitle

\begin{abstract}
Traditional speech emotion recognition (SER) evaluations have been performed
merely on a speaker-independent condition; some of them even did not evaluate
their result on this condition. This paper highlights the importance of
splitting training and test data for SER by script, known as sentence-open or
text-independent criteria. The results show that employing sentence-open
criteria degraded the performance of SER. This finding implies the difficulties
of recognizing emotion from speech in different linguistic information embedded
in acoustic information. Surprisingly, text-independent criteria consistently
performed worse than speaker+text-independent criteria. The full order of
difficulties for splitting criteria on SER performances from the most difficult
to the easiest is text-independent, speaker+text-independent,
speaker-independent, and speaker+text-dependent. The gap between
speaker+text-independent and text-independent was smaller than other criteria,
strengthening the difficulties of recognizing emotion from speech in different
sentences.

\end{abstract}
\section{Introduction}
\label{sec:intro}
Speech emotion recognition (SER) is one topic of interest in automatic speech
recognition and understanding. In contrast to automatic speech recognition (ASR)
which attempts to obtain linguistic information from speech, SER attempts to
obtain non-linguistic information from speech. In more concrete, SER aims to
infer the affective state of the speaker from solely speech data.

SER can be designed to recognize discrete emotion, continuous emotion, or both
emotion models. Recent research suggested that emotion is ordinal by nature
\cite{Yannakakis2021}, which is closer to a categorical than the continuous
model. In the categorical model, several emotion categories exist, from the
simplest two categories with positive and negative emotions to 27 categories
\cite{Cowen2017}. The choice of emotion model in SER depends on the availability
of the labels in the dataset.

Data-driven methods, in which most SER systems employ this kind of approach,
rely on the configuration or selection of the data to build the model. In SER,
it is common to split the data by evaluating different speakers for training and
test partitions. This approach, known as speaker-independent criteria, is a gold
standard to build SER model that minimizes speaker variability in the training
phase.

Recent research in SER, particularly by fusing acoustic and linguistic
information, has found that different splitting criteria in splitting data for
training will result in different performances \cite{Pepino2020}. In the fusion
of acoustic and linguistic information, it is sound that the model needs to be
trained in different scripts, known as sentence-open or text-independent. This
strategy was intended to avoid the effect of having the same linguistic
information for predicting emotion under the same sentences for both training
and test partitions. Since linguistic information is extracted via text or
script, this splitting condition is necessary to evaluate such discrepancies of
using different features or types of information. Using merely acoustic features
for SER, one may argue that this evaluation is unnecessary since no linguistic
features are involved in building the SER model.

Fujisaki in 2002 proposed a scheme in which various types of information are
manifested in the segmental and suprasegmental features of speech
\cite{Fujisaki2003}. One of the types of information includes emotion.
Referencing this argument that emotional information is manifested directly in
speech without a need to convert speech into text, there is a possibility that
different sentences will yield different SER performances under the same
acoustic-only system. The current research on SER showed no evaluation of the
differences of splitting criteria, particularly comparing data with and without
linguistic information.

The contribution of this paper is an evaluation of the effect of splitting
criteria into the training data on SER performance. As argued previously,
linguistic information is embedded in acoustic features; hence, evaluating
text-independent criteria, i.e., different sentences for training and test, is
necessary to observe such effects. We evaluated four splitting criteria:
speaker-dependent (including text-dependent data), speaker-independent,
text-independent, and speaker+text-independent criteria, and traced their SER
performances. We experimented with these criteria in three different
experiments. The results of three different experiments show a consistent
pattern of difficulties for four different splitting criteria. Text-independent
criteria obtained the worst result followed by speaker+text-independent,
speaker-independent, and speaker-dependent criteria.

\section{RELATED WORK}
\label{sec:format}
Research on the evaluation of the effect of data selection on speech processing
is not new. Different data for training will produce different results on the
same algorithm. A model generalization is a challenging task to minimize the
variance among different data inputs for the same model.

In \cite{Yamada2021}, the authors evaluated the effect of training data
selection for speech recognition of emotional speech. Emotional speech showed
lower error rates than neutral speech. Using their proposed selection criteria
based on the entropy of diphones, they improved the error rate of speech
recognition by transferring the model from emotional speech to non-emotional
speech data.

The authors of \cite{Pepino2020} pointed to the different results of splitting
data by random folds, speaker folds, and script folds on the speech emotion
recognition by fusing acoustic and linguistic information [Fig. 2]. The results
suggested that splitting by both speakers and scripts is more difficult than
splitting by random folds only or splitting by speaker only. No information is
available on comparing the difficulties of splitting by both speaker and script
to splitting by script only.

The author of \cite{Lee2019} evaluated Japanese Twitter-based emotional speech
(JTES) dataset with speaker-open and sentence-closed conditions. The author
achieved an accuracy of 81.44 \% by utilizing a common normalization. Using two
other corpora as training data and JTES as test data resulted in a degraded
score to 80.66\%, highlighting the difficulty of cross-corpus evaluation
\cite{Lee2021}. On the same JTES dataset but different criteria, speaker-open
and sentence-open, the authors of \cite{Chiba2020} achieved an accuracy of 73.4
\% by utilizing multi-stream attention-based bidirectional LSTM (BLSTM) with
feature segmentation. This result is the closest machine performance to human
performance on the JTES dataset, in which human evaluators scored 75.5\% from
acoustic subjective evaluation \cite{Mai2020}. However, again, there is no
evaluation comparing the effect of splitting by speakers (text-independent) and
other criteria.

This research fills the gap in the existing research on evaluating SER with
different splitting criteria. We evaluated speaker-independent (including
text-independent), speaker-dependent, text-independent, and
speaker+text-independent data. This paper denotes an insight to add information
missing in the previous papers to gain new insight on the effect of splitting
data based on the different criteria on SER performance.

\section{METHODS}

\begin{figure*}[htbp]
  \centering
  \includegraphics[width=\textwidth]{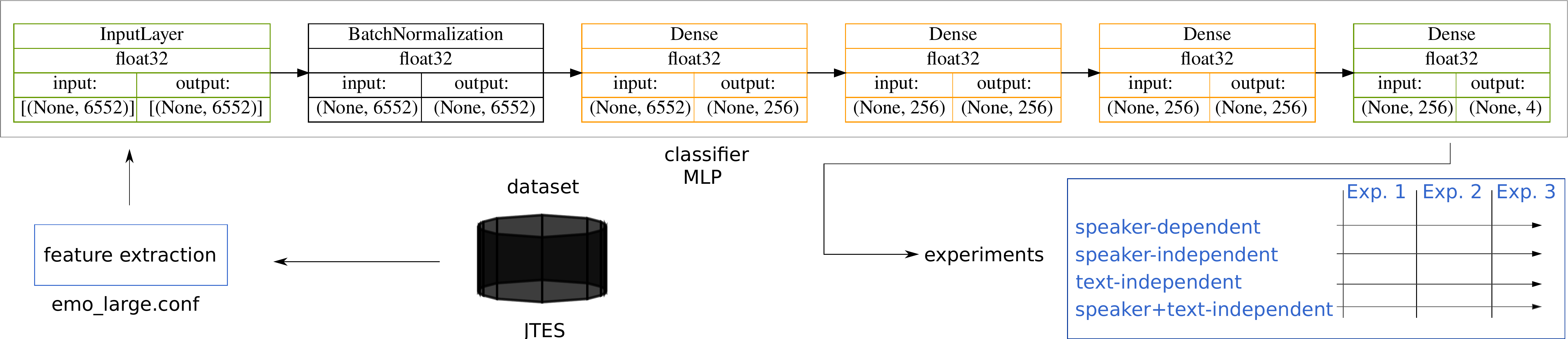}
  \caption{Flow diagram of research}
  \label{fig:flowdiag}
\end{figure*}

Fig. \ref{fig:flowdiag} shows the flow diagram of this research. Each component
on that diagram is explained below.

\label{sec:methods}

\subsection{DATASET}
\label{sec:pagestyle}
This paper makes use of the phonetically and prosodically balanced Japanese
Twitter-based Emotional Speech corpus (JTES) \cite{Takeishi2017}. The main
reason to choose this emotional speech dataset is due to the richness of
sentences (200 sentences: 50 sentences for each emotion) and speakers (100
speakers: 50 male and 50 female), which can be evaluated with different criteria
and conditions. Japanese speech emotion recognition also lacks exploration
compared to English, Chinese, and European languages.

The JTES dataset consists of 20000 utterances from four emotion categories: joy,
anger, sadness, and neutral. For each emotion, there are 500 raw sentences
collected from Twitter data. The Twitter text data are modified to reflect
Japanese culture when they are spoken in the recordings. For instance, sentences
that end with nouns were removed due to emotional independence. The labeling of
the emotion categories was performed by matching emotion-related words. 
Selection criteria based on entropy reduce the number into 50 sentences for each
emotion category.

All data in the JTES dataset were involved in the experiments. However, due to
the naturalness of this research's aims, each splitting criteria resulted in
a different number of data for each experiment. For instance, splitting by
speakers and by sentences will allocate 19600 samples for training, and the rest
400 samples for test. However, splitting by both speakers and sentences only
allocates 14400 samples for training due to overlap between both speaker and
text. To avoid the effect of different sizes of training data, we performed an
evaluation on the same amount of data.

\subsection{Feature Extraction}
The first step to obtaining input data for the SER system from the raw dataset
is by extracting features of speech. For this purpose, we utilized opensmile
feature extraction tool to extract a set of acoustic features that represent
emotion information in speech. The version of opensmile used in this experiment
is 3.0. The feature set is 'emo\_large' with default configuration
\verb|emo_large.conf| in the \verb|config/misc/| directory.

For this emo\_large feature set, we extracted only statistical functions per
utterance. The size was a 6552-dimensional feature from 56 low-level descriptors
with their deltas and deltas-deltas (total 168 features) multiplied by 39
statistics. For the feature data, first, we extracted each audio file (.wav)
into corresponding .csv files. Then we extracted related rows in .csv files
(rows 6559 to end) and saved them in the form of Numpy arrays (.npy files). The
first 6558 rows only consist of header names of the corresponding features,
which are not used in a classification process.

\begin{table}[htbp]
  \caption{Low-level descriptors (LLD) in emo\_large configuration}
  \centering\begin{tabular}{l | c}
    \hline
    Feat. number & LLD \\
    \hline
    1 & Log energy \\
    2-14 & MFCC (0-12) \\
    15-40 & Melspec (0-25) \\
    41 & Zero Crossing Rate (ZCR)\\
    42 & Voice probability \\
    44-44 & $f_o$, $f_o$ env \\
    45 & Spectral energy: 0-250 Hz \\
    46 & Spectral energy: 0-250 Hz \\
    47 & Spectral energy: 0-650 Hz \\
    48 & Spectral energy: 1000-4000 Hz \\
    49-52 & Spectral RollOff (25\%, 50\%, 75\%, 90\%)\\
    53-54 & Spectral Flux, Spectral Centroid \\
    55-56 & Spectral MaxPos, Spectral MinPos \\
    57-168 & $\Delta + \Delta \Delta $ \\
    \hline
  \end{tabular}
  \label{tab:emolarge}
\end{table}

Table \ref{tab:emolarge} shows detail of the acoustic features in emo\_large
setup. The top rows show fifty-six LLDs. Adding this 56 LLD with their deltas
($\Delta$) and deltas-deltas ($\Delta \Delta$) sums up 168 features. For each
LLD, 39 statistics were computed, resulting in 6652 features in total. Aside
from emotion recognition, this feature set is reported to be effective for
classifying dogs' barking \cite{Perez-Espinosa2019}.

\subsection{Classifiers}
We evaluated multi-layer perceptron (MLP), also known as fully-connected or
dense network, for the classifier of the SER system. The previous research on
speech emotion recognition has shown the effectiveness of these fully-connected
(FC) or dense networks to overcome the SER problem \cite{Atmaja2020k}. The
structure of FC networks follow the previous research on the different layers
and dataset \cite{Atmaja2020e}. Three dense layers were stacked, followed by a
four-unit dense layer as the final layer.

Table \ref{tab:dense} shows the detail of hyperparameters employed in the MLP
networks. The choice of the values is based on some references
\cite{Haneda2019, Atmaja2020e}. The same architecture was employed to evaluate
four different conditions in three experiments.

\begin{table}[htbp]
  \caption{Hyperparameters of the classifier}
  \centering\begin{tabular}{l c}
    \hline
    Parameter & Value \\
    \hline
    Networks/layers & FC \\
    Layer activation & ReLU \\
    Units   & (256, 256, 256) \\
    Optimizer & Adam \\
    Learning rate & 0.001 \\
    Epoch & 25 \\
    Batch size & 1024 \\
    Validation split & 20\% \\
    Output layer (units) & FC (4) \\
    Output activation & Softmax \\
    \hline
  \end{tabular}
  \label{tab:dense}
\end{table}

\subsection{Experiments}
\begin{figure}[htbp]
  \centering
  \includegraphics[width=0.45\textwidth]{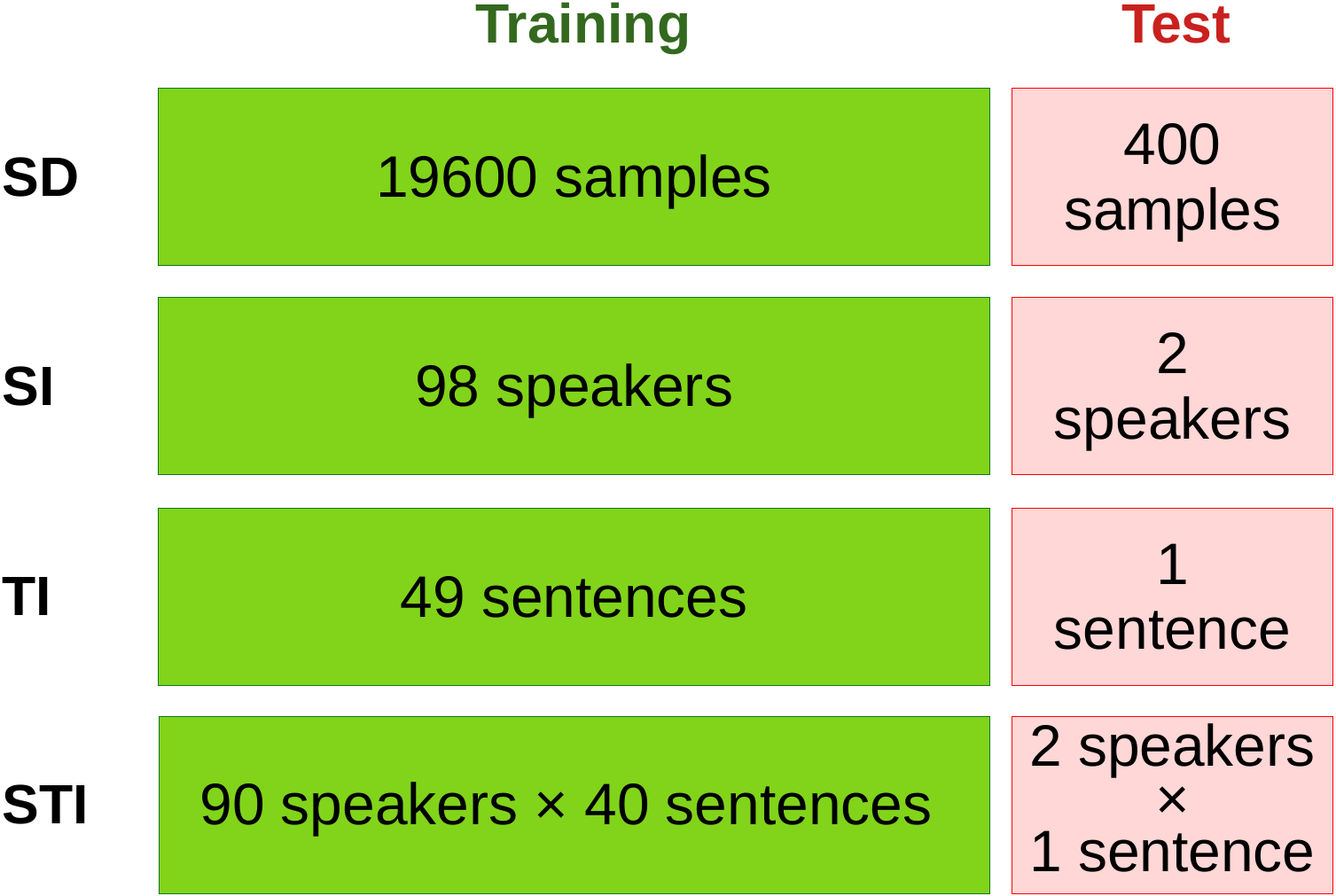}
  \caption{Data partition on each splitting criteria; the number of samples on
  each criteria is shown in Table \ref{tab:data}. }
  \label{fig:splitting}
\end{figure}

Since the goal of this research is to evaluate the performance of SER under
different training criteria, we split our experiment into four criteria, namely
speaker-dependent (SD), speaker-independent (SI), text-independent (TI), and
speaker+text-independent (STI). The splitting criteria were designed to match
the number of STI the test partition, i.e., 400 samples, following the previous
research \cite{Chiba2020,Haneda2019}. Fig. \ref{fig:splitting} depicts the
configuration of this splitting criteria. Note that the first SD criterion also
contains text-dependent data since the data are randomly shuffled. For instance,
we only selected the first 19600 utterances for training and the rest 400
utterances for the test for both first and second experiments. Both number of
data are obtained after shuffling the data. In the second, third, and fourth
criteria, we selected the samples such that the condition of each criterion is
fulfilled, i.e., the samples in the training and test partitions are different
based on these criteria. Three different experiments were conducted to gain an
overall conclusion among these splitting criteria.

In the first experiment, we ran 30 trials and calculated an average number of
accuracy (weighted accuracy, WA) along with its standard errors (SE). This
method is similar to the method reported in \cite{Chiba2020}. While they only
ran ten trials, we extended it into 30 trials for more statistical confidence.
This first experiment is conducted to attain an initial result and to compare
the trend with the later methods (cross-validation with same and different
number of data). The number of 30 trials was chosen since the results showed
unstable performance when the model was trained with less than this number of
trials.

The second experiment replaced an average score from a number of trials with
cross-validation. In this experiment, we divided each training data (SD, SI, TI,
STI) into five-folds and shuffle the training and validation data in different
five splits. The model generated by cross-validation with this 5-fold is then
used to predict the test data. As shown in Table \ref{tab:data}, both the first
and second experiments employed the same amount of data. For SD, SI, and TI,
19600 samples were used for training while the rest 400 samples were kept for
the test. For STI, which overlaps between speaker and text-independent data,
only 14400 can be used for training with the same number of test data, i.e., 400
samples.

To take into account the effect of different numbers of training data, we
reduced the number of training samples in the previous cross-validation
experiment. In this third experiment, all criteria have the same number of data
for both training and test partitions. A larger number of training data is
reported to be more effective for deep learning than smaller data
\cite{Halevy2009}. Hence, it is necessary to train the model under the same
number of data.

\begin{table}[htbp]
  \caption{Number of samples/utterances evaluated in different criteria for each experiment}
  \centering\begin{tabular}{l c c c c}
  \hline
  Criteria  & \multicolumn{2}{c}{Exp. \#1 \& Exp. \#2}	& \multicolumn{2}{c}{Exp. \#3}
  \\
  \hline
  {}          & Training	& Test	& Training	& Test \\
  SD, SI, TI	& 19600	& 400	& 14400	& 400 \\
  STI	        & 14400	& 400	& 14400	& 400 \\
  \hline
  \end{tabular}
  \label{tab:data}
\end{table}

\subsection{Evaluation Metric}
We evaluated our results, i.e., classification performance, with a single metric
weighted accuracy (WA). This metric is also known as overall accuracy
\cite{Satt2017}. Since the dataset is balanced, i.e., the number of
samples for each emotion class is same, various metrics (accuracy,
precision, recall, F1) result in similar performance. Hence, we only used WA as
a measure of total correct predicted samples divided by the number of all
samples irrespective of emotion classes. This metric is also known as weighted
average recall \cite[p. 268]{schuller2013computational}.

In addition to WA, we calculated error bars using the standard error (SE) of the
mean. This SE ``reflect the uncertainty in the mean and its dependency on the
sample size, $n$'' \cite{Krzywinski2013}. It can be formulated as,
\begin{equation}
  SE = \frac{SD}{\sqrt{n}},
\end{equation}
where SD is the standard deviation. As shown in the formula, SE depends on the
number of samples, $n$. In our experiment, $n = 30$ is for experiment 1 and
$n=5$ is for both experiment 2 and experiment 3.

\section{RESULTS AND DISCUSSION}
\label{sec:result}
\subsection{Results}
Table \ref{tab:result} shows the result of experiments under different
conditions. As noted in the Methods section, one of the purposes of conducting
experiments in different conditions (average of trials, cross-validation, and
same amount data) is to check the consistency and pattern of the evaluated
criteria. Hence, we summarized all results of these experiments in a single
table to show these behaviors.


Table \ref{tab:result} shows results of three SER experiment conditions in terms
of accuracy (weighted accuracy, WA). There is consistency among different
splitting criteria across different experiments. The SD condition always gains
the highest performance, as expected. Contrary to expectation, the
text-independent condition always obtains the lowest performance among others.
We first assumed that the speaker+text-independent condition would have the
worst result since this condition is the most difficult among others. These
results suggest that the order of difficulties across different splitting
criteria is text-independent, speaker+text-independent, speaker-independent, and
speaker-dependent.

\begin{table}[htbp]
  \caption{Results in weighted accuracy (WA) $\pm$ standard error (SE)
  on different experiment conditions; Exp. 1: average 30 trials; Exp. 2:
  cross-validation; Exp. 3: same amount data }
  \centering\begin{tabular}{l c c c}
    \hline
    Criteria  & \multicolumn{3}{c}{ WA (\%) $\pm$ SE (\%)}\\
    \hline
    {}  & Exp. \#1 & Exp. \#2 & Exp. \#3 \\
    SD	& 91.14	$\pm$ 0.07	& 92.30	$\pm$ 0.40	& 89.40	$\pm$ 0.48 \\
    SI	& 87.88	$\pm$ 0.09	& 88.85	$\pm$ 0.49	& 86.64	$\pm$ 0.63 \\
    TI	& 64.36	$\pm$ 0.08	& 65.04	$\pm$ 0.90	& 62.35	$\pm$ 0.93 \\
    STI	& 69.56	$\pm$ 0.09	& 70.65	$\pm$ 0.44	& 70.65	$\pm$ 0.44 \\
    \hline
  \end{tabular}
  \label{tab:result}
\end{table}

\subsection{Discussion}
We came to this research idea from the previous findings on SER research where
the recognition performance has increased when acoustic information is combined
with linguistic information \cite{Pepino2020,Atmaja2020k}. If linguistic
information is involved, it makes sense that the performance will improve. In
other words, the SER performance will decrease without linguistic information.
While the previous linguistic information is extracted from text, Fujisaki
\cite{Fujisaki2003} argued that linguistic information itself is also manifested
in acoustic features, meaning that there is also linguistic information in
acoustic features. This argument leads to the following presumption. If we train
SER with different splitting criteria with regard to linguistic information, the
performance also changes.

Implementing splitting criteria with different linguistic information can be
achieved by separating training and test sets into different sentences. The
spoken sentences (or utterances) used for the test partition can be set to be
different from the sentences used for the training. One may argue that this
separation only works if linguistic information from text is involved. However,
based on a solid argument from the previous literature \cite{Fujisaki2003}, we
believe that there will be differences in SER performance by splitting training
and test partitions into different sentences.

To test our hypothesis, we split the samples in the dataset into four criteria:
speaker-dependent, speaker-independent, text-independent,
speaker+text-independent. We experimented with these different splitting
criteria on three different conditions: average trials, cross-validation, and
the same amount of data. The results show consistency among different splitting
criteria and experiment conditions.

Among three experiments, the cross-validation condition obtained the highest
scores among the other two conditions. Shuffling different training and
validation data seems to be useful for generalizing the model. Using five-fold
shows slightly higher scores than an average of 30 trials. The results may also
be used to justify that reporting performance from average trials may be
sufficient to gain an insight into the order of difficulties among different
splitting criteria. One exception in guaranteeing this justification is that the
test set for four splitting criteria is different for various criteria in one
experiment condition. It is impossible to have the same test set for SD, SI, TI,
and STI. Indeed, the test sets are the same for each criterion across different
experiments.

The last experiment 3 using the same amount of data for all splitting criteria
seems to be the most relevant result in this research. As shown in a literature
\cite{Atmaja2019c}, the size of data greatly influences the performance of deep
learning, particularly on the SER task. To avoid bias due to differences in the
amount of training data, we forced all data in the third experiment to have the
same size. In this scenario, the results show consistency with the previous
experiments with smaller performance scores. The result for STI was the same for
both experiment 2 and experiment 3 since both experiments used the same data
for training and test.

One interesting finding in this research is that the text-independent criterion
is more difficult than speaker+text-independent criteria. A possible explanation
for this phenomenon might be that the model learns more information in
speaker+text-independent than in text-independent only. To test that hypothesis,
it is suggested for future research to explore the data inside the model using 
such a tool, e.g., t-SNE \cite{VanDerMaaten2008}. Testing the experimental setup
used in this research on other datasets might be useful to generalize the
finding, particularly on the order of difficulties across different splitting
criteria.

This research challenges the previously reported results on SER research, where
most evaluations are performed under speaker-independent only. One may argue
that building larger datasets that cover more spoken words may be sufficient to
tackle this limitation. However, in the currently evaluated datasets in SER
community, there are limited numbers of samples available. There is a necessity
to overcome this sentence-open problem in SER with such strategies.

\section{Conclusions}
\label{sec:conclu}
This paper reports experimental research on the effect of different splitting
criteria for speech emotion recognition tasks. Since linguistic information is
manifested in acoustic features, it is hypothesized that different splitting
criteria with regard to linguistic information will lead to different
performances on the SER task. Four splitting criteria were evaluated, focusing
on the differences between splitting criteria with the same and different
linguistic information. The first is known as text-dependent, while the latter
is known as text-independent. Along with splitting criteria by speakers, we
conducted experiments on three conditions. The results show a consistency that
text-independent condition is the most difficult condition among others.

To tackle the limitation of text-independent condition, one may utilize a
larger dataset to cover more spoken words for training SER. Under limited or
small dataset, it is necessary to find a strategy to learn information in
different sentences in training and test partitions.

\section{Acknowledgment}
This paper is based on results obtained from a project,
JPNP20006, commissioned by the New Energy and Industrial
Technology Development Organization (NEDO), Japan.

\bibliographystyle{IEEEtran}
\bibliography{jtes}

\end{document}